\documentclass[twocolumn,prl,showpacs,amsmath,amssymb,floatfix]{revtex4}
\usepackage{graphicx}
\usepackage{dcolumn}
\usepackage{bm}

\begin{document}

\title{Percolation and Loop Statistics in Complex Networks}

\author{Jae Dong Noh}
\affiliation{Department of Physics, University of Seoul,
  Seoul 130-743, Korea}

\date{\today}
\begin{abstract}
Complex networks display various types of percolation transitions. We show
that the degree distribution and the degree-degree correlation alone are not
sufficient to describe diverse percolation critical phenomena.
This suggests that a genuine structural correlation is an essential
ingredient in characterizing networks.
As a signature of the correlation we investigate a scaling behavior in
$M_N(h)$, the number of finite loops of size $h$, with respect to 
a network size $N$.  
We find that networks, whose degree distributions are
not too broad, fall into two classes exhibiting 
$M_N(h)\sim (\mbox{constant})$ and $M_N(h) \sim (\ln N)^\psi$, respectively.
This classification coincides with the one according to the percolation
critical phenomena.
\end{abstract}
\maketitle

{\bf Introduction:} 
Complex networks have been attracting much interest during the last decade.
Structure, dynamics, and collective phenomena have become intriguing
research subjects in network 
science~\cite{Watts98,Albert02,Boccaletti06,Dorogovtsev07}.  
This work considers a structural correlation in complex networks.
Discussing scaling behaviors of percolation
transitions~\cite{Callaway00,Cohen02,Lee04,Newman02,Vazquez03,Callaway01,Dorogovtsev01,JKim02,Krapivsky04,Noh07}, 
we will show that the structural correlation is an essential ingredient
characterizing structural properties and collective phenomena
of complex networks.

Structural inhomogeneity is one of the most salient features of complex
networks. It is reflected on the broad degree distribution 
$p_d$~\cite{Albert02}. 
For the study of the inhomogeneity, the {\em uncorrelated} 
network has been considered. It is defined as an ensemble of networks
specified only with a given degree distribution but random in any
other aspect.  This includes the Erd\H{o}s-R\'enyi random
graph~\cite{Albert02},
the Molloy-Reed model~\cite{Molloy95}, and the static
model~\cite{Goh01,Lee06}. 
It is revealed that the structural inhomogeneity encoded in $p_d$
leads to rich collective phenomena~\cite{Dorogovtsev07}.

Many real-world networks display some structural correlations. 
The degree-degree~(DD) correlation has been considered 
mostly~\cite{Pastor-Satorras01,Newman02}. 
It refers to the correlation between degrees of nodes 
linked directly with an edge. 
Networks with a positive~(negative) correlation
are called to be assortative~(disassortative).
The DD correlation is represented with the 
two-point degree correlation function $p(d',d)$ denoting the fraction of edges 
linking nodes of degree $d$ and $d'$.  
With the DD correlation being incorporated, the {\em randomly
correlated}~(RC) network has been considered.
It is defined as an ensemble of networks specified only with 
$p(d',d)$, but again random 
in any other aspect~\cite{Newman02,Vazquez03,Dorogovtsev04}.

Studying the RC network is a meaningful attempt 
to understand correlated networks.
However, it remains unknown to what extent the RC
network is a proper model for correlated networks.
In this work we will show that there exists a class of networks with a
genuine structural correlation that cannot be captured by the DD correlation. 
This will be shown by discussing percolation critical phenomena in 
various networks.
We will also suggest that number statistics 
of loops is useful as a signature of the correlation.

{\bf Percolation:} Consider a bond percolation problem with 
the occupation probability $f$ in the RC network. It is specified 
with a degree correlation function $p(d',d)$. 
Other characteristics are easily represented with it~\cite{Newman02}. 
For examples, $q_d = \sum_{d'} p(d',d)$ is the probability 
that an edge chosen randomly 
leads to a node of degree $d$. It is related to the degree
distribution as $q_d = d p_d / \langle d\rangle$ with the mean degree
$\langle d \rangle$. 
The conditional probability that the degree of a neighbor of a degree-$d$ node is $d'$
is given by $p(d'|d) = p(d',d)/q_d$.
The uncorrelated network has 
the property that $p(d',d) = q_{d'} q_{d}$ or $p(d'|d) = q_{d'}$.

We summarize briefly the theory for the percolation in the RC 
network~(see Refs.~\cite{Newman02,Vazquez03} for detail).
The percolation order parameter $P_\infty$, the fraction 
of nodes in the infinite cluster, is given by
\begin{equation}\label{Pinfty}
P_\infty = 1 - \sum_d p_d \left(u_d\right)^d \ ,
\end{equation}
where $\{u_d\}$ should satisfy the self-consistent equation
\begin{equation}\label{sc_for_u}
u_d = (1-f) + f \sum_{d'} p(d'|d) \left(u_{d'}\right)^{d'-1}  \ .
\end{equation}
Here $u_d$ denotes the probability that an edge 
from a degree-$d$ node leads to a finite cluster. 
It has a trivial solution $\{u_d=1\}$ corresponding to $P_\infty=0$.
Hence the percolation threshold can be
obtained from the linear stability analysis around the trivial
solution.
It leads to $ f_c = 1 / \Lambda_{max}$~\cite{Newman02,Vazquez03},
where $\Lambda_{max}$ is the maximum
eigenvalue of the matrix ${\bf C}$ with elements
\begin{equation}\label{Cdd}
C_{dd'} = (d'-1) p(d' | d) \ .
\end{equation}
For the uncorrelated network having $p(d'|d) = q_{d'}$,
it yields the well-known result $f_c = \langle d \rangle /
(\langle d^2\rangle - \langle d\rangle)$~\cite{Cohen02}.

One can also find the mean cluster size $S$, average size of finite
clusters enclosing a node chosen randomly~\cite{Newman02,Vazquez03}.
For the sake of simplicity, we consider only
the region $f<f_c$ where $u_d = 1$ 
for all $d$. 
The same conclusion can be drawn in the region
$f>f_c$, which is not presented here.
It is given by 
\begin{equation}\label{S}
S = 1 + \sum_d \ d ~ p_d ~ v_d \ ,
\end{equation}
where $\{v_d\}$ should satisfy the self-consistent equation
$ v_d = f + f \sum_{d'} C_{d d'}  v_{d'}$. 
Using the matrix and vector notation, 
the solution is given by
\begin{equation}
\vec{v} = \left( {\bf I} - f {\bf C} \right)^{-1} \cdot \vec{f} \ ,
\end{equation}
where $\vec{v} = (v_1,v_2,\cdots)$, $\vec{f} = (f,f,\cdots)$, and  
${\bf I}$ is the identity matrix. 

The onset of the percolation transition and the mean cluster
size is determined with the matrix $\mathbf{C}$.
As one approaches the percolation threshold $f_c=1/\Lambda_{max}$, 
the largest eigenvalue of
$f{\mathbf C}$ approaches unity, the matrix 
$\left({\bf I} - f {\bf C}\right)^{-1}$ becomes singular, 
and $v_d$'s become divergent~\cite{Newman02}. 
Consequently, the mean cluster size $S$ diverges as $f\rightarrow f_c$.
This analysis shows that the percolation in the RC network
as well as in the uncorrelated network should be accompanied with 
divergent $S$ regardless of the shape of $p(d',d)$. 

On the other hand, recent studies have revealed that some networks 
display percolation transitions which are associated with 
non-divergent mean cluster sizes.
These are the growing network~(GN)
models~\cite{Callaway01,Dorogovtsev01,JKim02,Krapivsky04} and 
the exponential random graph~(ERG) model~\cite{Noh07}. 
The GN model is defined through a growth rule. In the GN model of
Callaway {\em et al.}~\cite{Callaway01}, for instance, a node is added each
time step and an edge is added with the probability $\delta$ between a 
randomly selected pair of nodes. 
This model undergoes an infinite order percolation 
transition at $\delta_c=1/8$ and 
the mean cluster size $S$ remains {\em finite} at the transition point. 
The degree correlation function $p(d',d)$ of the network is
known~\cite{Callaway01}. With the correlation function,
one might approximate the network as the RC network and apply
the above theory.
However, it would yields that the mean cluster size diverges as
$S \sim |\delta-\delta_c|^{-1.0}$ with $\delta_c\simeq
0.192$~\cite{Noh_unpub}, which is not the case.

Another example is the ERG~\cite{Noh07} which is defined as the equilibrium
ensemble of networks~\cite{Park04} with the Poisson degree distribution.
The model has an interesting feature that one can adjust the strength of
the DD correlation with a parameter 
$J$~(see Ref.~\cite{Noh07} for details).
The network is assortative~(disassortative) 
with positive~(negative) values of $J$, and uncorrelated when $J=0$. 
Numerical study of the percolation
in Ref.~\cite{Noh07} showed that the disassortative network 
with $J=-1$ belongs to the same universality class as the uncorrelated 
network with $J=0$.
On the contrary, the assortative network with $J=1$ was found to display
the similar type of percolation transition to the GN model.
That is to say, the mean cluster size $S$ is finite at the 
percolation threshold.

A common feature of the GN models and the ERG model with positive $J$ 
is an assortative DD correlation.
However, it is evident that they cannot be described as the RC network.
It suggests that there exists a genuine structural correlation 
that is responsible for the distinct universality class of 
the percolation transitions and cannot be captured only by the DD correlation. 

We propose that the correlation can be characterized with 
loop structure.
An $h$-loop or $h$-cycle is defined as a self-avoiding path through 
$h$ distinct nodes. 
Loops in complex networks have been studied in 
literatures~\cite{Watts98,Dorogovtsev04,zrp_noh,rw_noh,Bianconi03,Lee04,Marinari04,Bianconi05,Rozenfeld05,Bianconi06}.
The clustering is related to the number of triangles, loops of size
$h=3$~\cite{Watts98,Dorogovtsev04}.
Dynamic scaling behaviors on networks are shown to depend on
whether there are loops or not~\cite{zrp_noh,rw_noh}.
Loops of system sizes are also 
studied~\cite{Marinari04,Bianconi05,Bianconi06}.
A loop forms when two end nodes of a linear path are 
linked to each other. 
In an uncorrelated sparse network of size $N$, such an event may occur 
by chance with the probability of the order of ${\cal O}(1/N)$.
Any structural correlation may be detected from deviations from 
uncorrelated network results.

{\bf Loops in uncorrelated and RC networks:}
We first consider, as an uncorrelated network, 
the static model~\cite{Goh01} with $N$ nodes and $KN$ edges. 
Each node $i=1,2,\cdots,N$ is assigned to a selection probability 
$\omega_i = i^{-\mu} / \left(\sum_{j=1}^N j^{-\mu}\right)$ with a parameter 
$0\leq \mu<1$. Edges are added successively by connecting two nodes chosen
with the selection probability. This leads to a scale-free network with 
the degree exponent $\gamma=1+1/\mu$, which
is free from the DD degree correlation for $\mu<1/2$ or
$\gamma>3$~\cite{Lee06}.

In evaluating the number of loops, it is crucial to find
the connecting probability $f_{ij}$ between two nodes $i$ and $j$.
The mean number $M_N(h)$ of $h$-loops is then given by
$ M_N(h) = \frac{1}{(2h)} \sum_{i_1,\cdots,i_h} 
f_{i_1i_2} f_{i_2 i_3}\cdots f_{i_h i_1}$
with the factor $1/(2h)$ compensating for overcounting.
In the sum the indices should be mutually distinct,
which will be neglected because it leads to a sub-leading order
correction.

The connecting probability is given by 
$f_{ij} = 1 - e^{-2KN\omega_i\omega_j}$~\cite{Lee06}. 
For $\mu<1/2$ or $\gamma>3$, one can use the approximation 
$f_{ij}  \simeq 2KN \omega_i \omega_j$~\cite{Lee06}.
This yields that
\begin{equation}\label{mh2_static} 
M_N(h) = \frac{(2KN)^h}{2h} I^h 
\end{equation}
with $I\equiv \sum_{i=1}^N \omega_i^2$. In the large $N$ limit,
it is given by $I = \frac{(1-\mu)^2}{ (1-2\mu) } N^{-1}$.
Inserting this into Eq.~(\ref{mh2_static}), we obtain
\begin{equation}
M_N(h) = \frac{1}{2h} \left( \frac{2K(1-\mu)^2}{1-2\mu} \right)^h  \ .
\end{equation}
The result shows that $M_N(h)$ is finite for any finite $h$ when $\gamma>3$.
In the case when $\mu > 1/2$ or $\gamma<3$, 
$M_N(h)$ grows algebraically with $N$~\cite{Dkim07}.

We also consider the RC network specified with a degree
correlation function $p(d',d)$. The number of $3$-loops or triangles 
in the RC network was studied in Ref.~\cite{Dorogovtsev04}.
One can generalize the analysis to find $M_N(h)$ for arbitrary finite values
of $h$. It is given by~\cite{Noh_unpub}
\begin{equation}
M_N(h) = \frac{1}{2h} \sum_{d_1,\cdots,d_h} \prod_{l=1}^h \left[ (d_l -1)
p(d_{l+1}|d_l) \right] \ ,
\end{equation}
where $d_{h+1}\equiv d_1$. In the uncorrelated limit where
$p(d|d')=q_d=dp_d/\langle d\rangle$, it reduces to $M_N(h)=( (\langle
d^2\rangle - \langle d\rangle)/\langle d\rangle)^h/(2h)$ which coincides
with the one obtained for the Molloy-Reed model~\cite{Bianconi05}.
Analyzing this formula, we find that $M_N(h)$ is also finite 
in the RC network unless the
DD correlation is so perfect that $p(d|d')$ is peaked at $d=d'$. 
Detailed analysis will be presented elsewhere~\cite{Noh_unpub}.

{\bf Loops in correlated networks:} 
First we consider the GN model of 
Callaway {\em et al.}~\cite{Callaway01}. 
A node added at $i$th step will be labeled with the index $i=1,\cdots,N$.
The growth rule implies that the
connecting probability $f_{ij}$ for $i>j$ is given by
$$
f_{ij} = 1-\prod_{n=i+1}^N \left( 1 - \frac{\delta}{ {}_n C_2 } \right)
\simeq \frac{2\delta}{N} \frac{ N - i}{i} + {\cal O}(N^{-3}) \ .
$$
It is not symmetric in $i$ and $j$, for which 
one needs some caution in evaluating $M_N(h)$.

\begin{figure}[t]
\includegraphics[width=\columnwidth]{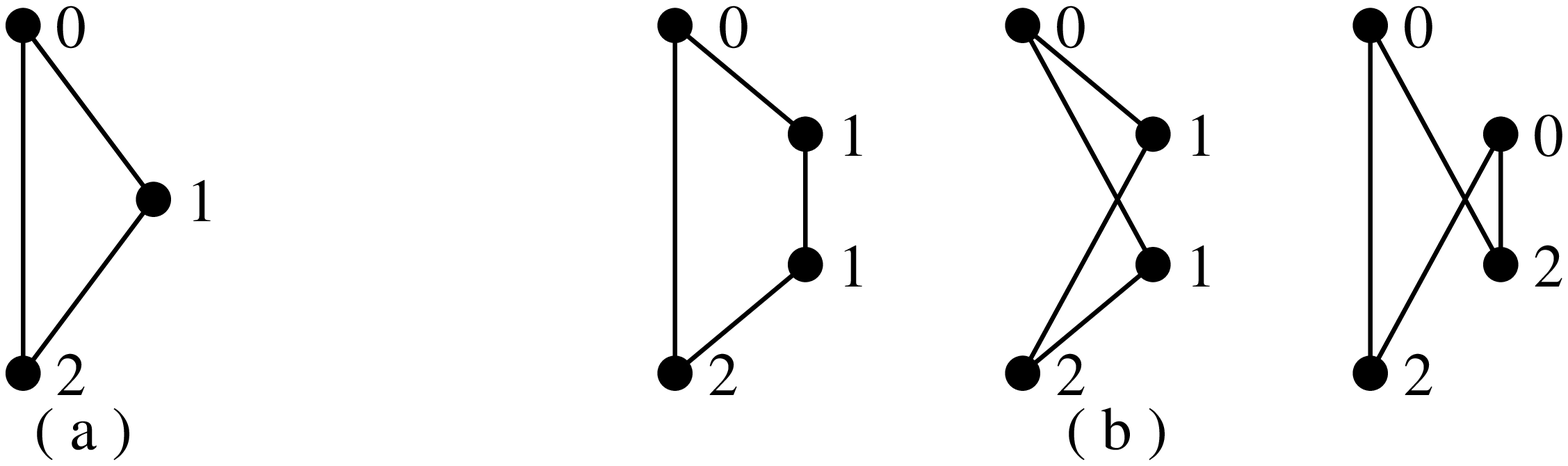}
\caption{Loop configurations of nodes $i_1 > \cdots >i_h$ from bottom to top
with $h=3$ in (a) and $h=4$ in (b). 
Also shown are the indices $\alpha_n$~(see text).}
\label{fig1}
\end{figure}

For three nodes $i_1>i_2>i_3$, a possible 3-loop configuration is unique as
shown in Fig.~\ref{fig1}(a). So one has that 
$M_N(3) = \sum_{i_1>i_2>i_3} f_{i_1 i_2} f_{i_2 i_3}
f_{i_3 i_1}$. In the large $N$ limit, 
it can be written as
\begin{equation}
M_N(3) = (2\delta)^3 \int [{\cal D}_3 {\mathbf x}]\  
g(x_2) \left(g(x_1)\right)^2 \ ,
\end{equation}
where $\int[{\cal D}_h {\mathbf x}]\equiv
\int_{1/N}^1 dx_h \int_{x_h}^1 dx_{h-1} \cdots \int_{x_2}^1 dx_1$
and  $g(x) \equiv {(1-x)}/{x}$.
Note that the integral is dominated by the contribution near $x_n \simeq 0$.
So, hereafter we will approximate the function $g(x)=(1-x)/x$ as $g(x)=1/x$,
which does not change the leading order behavior.
Evaluating the integral, we obtain that
\begin{equation}
M_N(3) = (2\delta)^3 \ln N \ .
\end{equation}

For four nodes $i_1>i_2>i_3>i_4$, there are three distinct loop
configurations as shown in Fig.~\ref{fig1}(b). 
One can count the number of loops in each configuration separately.
Summing them up, one obtains that
$ M_N(4) = (2\delta)^4 \int[{\cal D}_4 {\mathbf x}] \left( 
2 g(x_3) g(x_2) \left(g(x_1)\right)^2 + \left(g(x_2)\right)^2 \left( g(x_1)
\right)^2 \right)$.
It yields that
\begin{equation}
M_N(4) = \frac{5}{2} (2\delta)^4 \ln N \ .
\end{equation}

%%%%% figure 1 %%%%%
%%%%% figure 1 %%%%%

For $h$ nodes $i_1 > \cdots > i_h$, the number of loop
configurations is given by $h!/(2h)$.
A loop configuration $\mathcal{C}$ can be specified with
$\alpha_n$~($\beta_n$), defined as
the number of older~(younger) partners of node $i_n$~(see Fig.~\ref{fig1}). 
They are non-negative and satisfy 
$\alpha_n + \beta_n = 2$ for all $n$.
For later use, we also define the quantities
$A_n \equiv \sum_{l=1}^n \alpha_l$ and $B_n \equiv \sum_{l=1}^n \beta_l$ 
satisfying $A_n + B_n = 2n$. 
Then the number of $h$-loops of configuration $\mathcal{C}$
is given by $M_N^{\mathcal{C}} (h) = (2\delta)^h 
\int[\mathcal{D}_h \mathbf{x}] \prod_{n=1}^h 
\left(g(x_n)\right)^{\alpha_n}$.
It yields that
\begin{equation}
M_N^{\mathcal{C}}(h) =  
(2\delta)^h \left[ \prod_{n=1}^{h-1} \frac{1}{ A_n  - n}\right] \ln N .
\label{mhC}
\end{equation}
It was assumed that $A_n\neq n$ for $n<h$, which can be proved 
easily~\cite{comment3}.
Summing over all $\mathcal{C}$, we obtain that
\begin{equation}\label{mh_callaway}
M_N(h) = C_h (2\delta)^h \ln N \ ,
\end{equation}
where the combinatorial factor $C_h$ takes the value $C_1=1$, $C_2=5/2$,
$C_3 = 7$, etc~\cite{comment4}. 
In contrast to the uncorrelated and the RC networks, 
the total number of $h$-loops 
in the GN model shows the logarithmic scaling.

We also study the correlated scale-free networks. 
Dorogovtsev {\em et al.}~\cite{Dorogovtsev01} 
extended the GN model of Callaway {\em et al.}~\cite{Callaway01} 
by adopting the preferential selection rule;
upon creating edges, 
a node with degree $d$ is selected with the probability proportional 
to $(d+a)$ with a parameter $a$.
The resulting networks are scale-free with the degree
exponent $\gamma=2+a/(2\delta)$. It is straightforward to show that the
connecting probability $f_{ij} \propto i^{-(\gamma-2)/(\gamma-1)}
j^{-1/(\gamma-1)}$ for two nodes $i>j$. It is asymmetric 
as in the previous case. So one can follow the same procedure to
calculate $M_N(h)$. We also find the logarithmic scaling $M_N(h) \sim \ln N$ 
for $\gamma>3$. Details will be presented elsewhere~\cite{Noh_unpub}.

\begin{figure}[t]
\includegraphics*[width=\columnwidth]{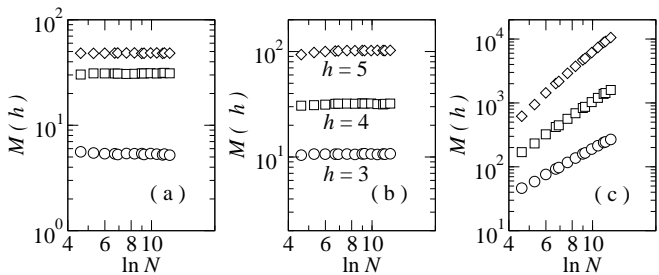}
\caption{Numbers of $h$-loops in the ERP model with a disassortative~(a),
neutral~(b), and assortative~(c) degree-degree correlation. }
\label{fig2}
\end{figure}

Finally we study numerically the loop statistics in the ERG 
model~\cite{Noh07}. An ensemble of ERG networks is generated using 
the Monte Carlo method explained in Ref.~\cite{Noh07}
and the number of loops is enumerated numerically.
Figure~\ref{fig2} presents the numerical results for $M_N(h)$ in the networks
with a disassortative~($J=-1$), the neutral~($J=0$), 
and an assortative~($J=1$) DD correlation, respectively.
We find that $M_N(h)$'s are finite in the disassortative network as well
as in the neutral network. 
It is noteworthy that both networks
display the percolation transition in the same universality
class~\cite{Noh07}.
On the contrary, we find a logarithmic scaling $M_N(h) 
\sim (\ln N)^{\psi_h}$ for the assortative network. 
This suggests that the assortative DD correlation, although insufficient, 
may give rise 
to the structural correlation that allows one to distinguish them
networks from others.
We remark that the exponent $\psi_h$ seems to depend on $h$ in the
assortative ERG model while it is independent of $h$ 
in the other correlated networks. 
Its reason and implication has not been understood yet.

%%%%% figure 2 %%%%%
%%%%% figure 2 %%%%%

{\bf Summary:} 
We have studied the number of loops in complex networks. 
Discussing the scaling behaviors of percolation critical phenomena, we
have shown that the structural correlation is a relevant
feature of networks.
The scaling behavior in the number $M_N(h)$ of $h$-loops for finite $h$ 
with network size $N$ is suggested as a signature of the correlation. 
When the degree distribution is not too broad,
we have found that $M_N(h)$ is finite and independent of $N$ in
one class of networks and that $M_N(h)$ displays the logarithmic scaling
$M_N(h)\sim (\ln N)^{\psi_h}$ in the other class of networks. 
The former includes the uncorrelated and the RC networks. The percolation
transitions in those networks are characterized with the divergent mean 
cluster size.
The latter includes the GN models and the assortative ERG model. 
The display the percolation transitions with
non-divergent mean cluster size. They have the assortative DD
correlation commonly, although the assortativity by itself is not a sufficient
condition.
Our work shows that the structural correlation, reflected 
on the loop statistics, is important for the percolation critical phenomena. 
Relevance to dynamical and equilibrium critical phenomena will 
be studied later.

Acknowledgement: This work was supported by Korea Research Foundation Grant
funded by the Korean Government (MOEHRD, Basic Research Promotion Fund)
(KRF-2006-003-C00122). The author thanks Doochul Kim, Hyunggyu Park, 
Byungnam Kahng, and Sergey Dorogovtsev for helpful discussions.


\begin{thebibliography}{50}
\bibitem{Watts98} D.J. Watts and S.H. Strogatz, 
        Nature (London) {\bf 393}, 440 (1998).
\bibitem{Albert02} R. Albert and A.-L. Barab\'asi, 
        Rev. Mod. Phys. {\bf 74}, 47 (2002).
\bibitem{Boccaletti06} S. Baccaletti, V. Latora, Y. Moreno, M. Chavez, 
        and D.-U. Hwang, Phys. Rep. {\bf 424}, 175 (2006).
\bibitem{Dorogovtsev07} S.N. Dorogovtsev, A.V. Goltsev, and J.F.F. Mendes, 
        arXiv:0705.0010

\bibitem{Callaway00} D.S. Callaway, M.E.J. Newman, S.H. Strogatz, 
        and D.J. Watts, Phys. Rev. Lett. {\bf 85}, 5468 (2000).
\bibitem{Cohen02} R. Cohen, D. ben-Avraham, and S. Havlin, 
        Phys. Rev. E {\bf 66}, 036113 (2002).
\bibitem{Lee04} D.-S. Lee, K.-I. Goh, B. Kahng, and D. Kim, 
        Nucl. Phys. B {\bf 696}, 351 (2004).

\bibitem{Newman02} M.E.J. Newman, Phys. Rev. Lett. {\bf 89}, 208701 (2002);
        Phys. Rev. E {\bf 67}, 026126 (2003).
\bibitem{Vazquez03} A. V\'azquez and Y. Moreno, 
        Phys. Rev. E {\bf 67} 015101(R) (2003).

\bibitem{Callaway01} D.S. Callaway, J.E. Hopcroft, J.M. Kleinberg, M.E.J.
Newman, and S.H. Strogatz, Phys. Rev. E {\bf 64}, 041902 (2001).
\bibitem{Dorogovtsev01} S.N. Dorogovtsev, J.F.F. Mendes, and A.N. Samukhin,
        Phys. Rev. E {\bf 64}, 066110 (2001).
\bibitem{JKim02} J. Kim, P.L. Krapivsky, B. Kahng, and S. Redner,
        Phys. Rev. E {\bf 66}, 055101(R) (2002).
\bibitem{Krapivsky04} P.L. Krapivsky and B. Derrida,
        Physica A {\bf 340},714 (2004).

\bibitem{Noh07} J.D. Noh, arXiv:0705.0087

\bibitem{Molloy95} M. Molloy and B. Reed, 
        Random Struct. Algorithms {\bf 6}, 161 (1995).
\bibitem{Goh01} K.-I. Goh, B. Kahng, and D. Kim,
        Phys. Rev. Lett. {\bf 87}, 278701 (2001).
\bibitem{Lee06} J.-S. Lee, K.-I. Goh, B. Kahng, and D. Kim,
        Eur. Phys. J. B {\bf 49}, 231 (2006).

\bibitem{Pastor-Satorras01} R. Pastor-Satorras, A. V\'azquez, and A.
Vespignani, Phys. Rev. Lett. {\bf 87}, 258701 (2001).

\bibitem{Dorogovtsev04} S.N. Dorogovtsev, 
         Phys. Rev. E {\bf 69}, 027104 (2004).

\bibitem{Noh_unpub} J.D. Noh, unpublished.

\bibitem{Park04} J. Park and M.E.J. Newman, 
        Phys. Rev. E {\bf 70}, 066117 (2004).

\bibitem{zrp_noh} J.D. Noh, G.M. Shim, and H. Lee, 
         Phys. Rev. Lett. {\bf 94}, 198701 (2005);
         J.D. Noh, Phys. Rev. E {\bf 72}, 056123 (2005).
\bibitem{rw_noh} J.D. Noh and S.-W. Kim, 
         J. Korean Phys. Soc. {\bf 48}, S202 (2006).

\bibitem{Bianconi03} G. Bianconi and A. Capocci,
        Phys. Rev. Lett. {\bf 90}, 078701 (2003).
\bibitem{Rozenfeld05} H.D. Rosenfeld, J.E. Kirk, E.M. Bolt, 
         and D.  ben-Avraham, J. Phys. A {\bf 38}, 4589 (2005).
\bibitem{Marinari04} E. Marinari and R. Monasson,
        J. Stat. Mech.: Theory Exp. P09004 (2004).
\bibitem{Bianconi05} G. Bianconi and M. Marsili,
        J. Stat. Mech.: Theory Exp. P06005 (2005).
\bibitem{Bianconi06} G. Bianconi and M. Marsili,
        Phys. Rev. E {\bf 73}, 066127 (2006).

\bibitem{Dkim07} D. Kim, in private communication.

\bibitem{comment3} 
To a given value of $n<h$,
one can separate the edges into three nonempty sets; 
one for edges among $\{i_1,\cdots,i_n\}$, 
another for edges among $\{i_{n+1},\cdots,i_h\}$, 
and the other for edges interconnecting them. 
An edge in the first set adds unity to $A_n$ and $B_n$, while one in 
the second set does not contribute to them. 
One the other hand, a edge in the third set adds 
unity to $A_n$ but nothing to $B_n$. 
Hence one finds that $A_n \geq B_n$ for all $n$. Since $A_n+B_n=2n$, it
proves that $A_n\geq n$.  The equality holds only when $n=h$. 

\bibitem{comment4} A closed form expression for $C_h$ is not known. 
An exact enumeration study suggests that it grows exponentially as
$C_h \sim c^h$ with $c\simeq 4.0$. 

\end{thebibliography}
\end{document}